\documentclass[aps,pre,reprint,unsortedaddress,showpacs,amsmath,amssymb,twocolumn,preprintnumbers]{revtex4-1}

\usepackage{graphics,graphicx}

\begin{document}
\preprint{Physical Review E}

\title{Mathematical model for logarithmic scaling of velocity fluctuations in wall turbulence}

\author{Hideaki Mouri}
\affiliation{Meteorological Research Institute, Nagamine, Tsukuba 305-0052, Japan}



\begin{abstract}
For wall turbulence, moments of velocity fluctuations are known to be logarithmic functions of the height from the wall. This logarithmic scaling is due to the existence of a characteristic velocity and to the nonexistence of any characteristic height in the range of the scaling. By using mathematics of random variables, we obtain its necessary and sufficient conditions. They are compared with characteristics of a phenomenological model of eddies attached to the wall and also with those of the logarithmic scaling of the mean velocity.

\end{abstract}


\maketitle

\section{Introduction} \label{S1}

\begin{figure}[bp]
\rotatebox{90}{
\resizebox{4.8cm}{!}{\includegraphics*[4.4cm,3.5cm][17.5cm,26.4cm]{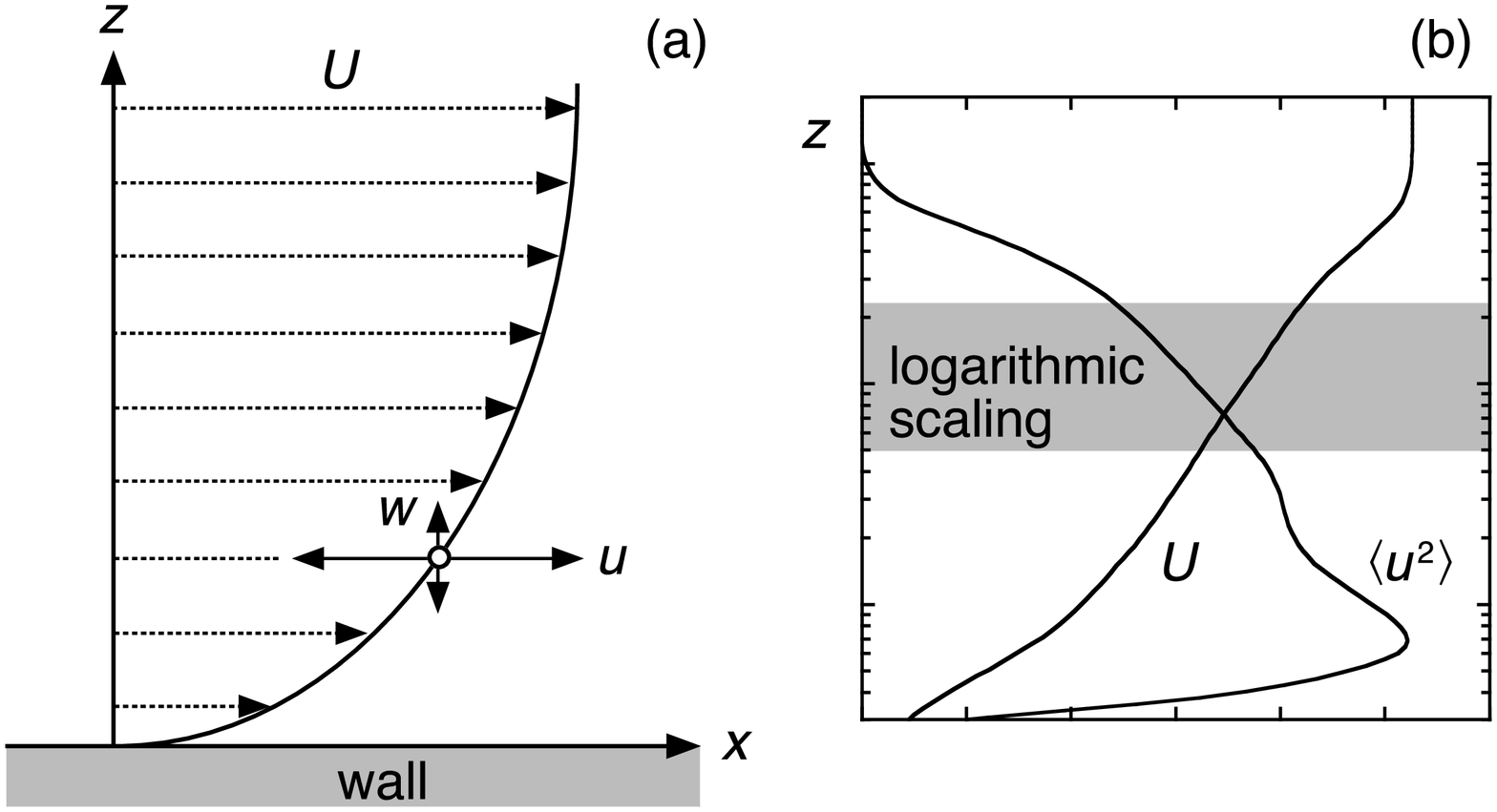}}
}
\caption{\label{f1} Wall turbulence over a flat plate. (a) Velocity components $U_z+u_z$ and $w_z$. (b) Profiles of $U_z$ and $\langle u_z^2 \rangle$ against the height $z$.}
\end{figure}

Consider wall turbulence in a cylindrical pipe, in a rectangular channel, or over a flat plate. If it is stationary and if its Reynolds number is high enough, it has a layer with a constant value of the mean rate of the momentum transfer, i.e., of the Reynolds stress $\rho \langle -u_z w_z \rangle$. Here $\rho$ is the mass density, $u_z$ and $w_z$ are velocity fluctuations in the streamwise and vertical directions at the height $z$ from the wall surface at the $x$-$y$ plane (see Fig.~\ref{f1}), and $\langle \cdot \rangle$ is the time average.

This constant-stress layer is known to exhibit a logarithmic scaling of the mean streamwise velocity $U_z$, which increases with an increase in the height $z$ as
\begin{equation}
\label{eq1}
\frac{U_{z_1}-U_{z_2}}{u_{\ast}} = \frac{1}{\kappa} \ln \left( \frac{z_1}{z_2} \right) .
\end{equation}
Here $u_{\ast} = \langle -u_z w_z \rangle^{1/2}$ is the friction velocity. The von K\'arm\'an constant $\kappa$ is about $0.4$, which is considered as a universal value \cite{mmhs13}. Since $U_z$ depends also on the outside of the layer, we focus on its difference $U_{z_1}-U_{z_2}$.

Recent experiments and simulations have revealed the existence of another logarithmic scaling for even-order moments of the streamwise velocity fluctuations $u_z$ \cite{hvbs12,mmhs13,mm13,hvbs13,swm14,vhm15}. They decrease with an increase in the height $z$ as

\begin{subequations}
\label{eq2}
\begin{equation}
\label{eq2a}
\frac{\langle u_{z_1}^{2m} \rangle^{1/m} -\langle u_{z_2}^{2m} \rangle^{1/m}}{u_{\ast}^2} 
= -\alpha_m \ln \left( \frac{z_1}{z_2} \right) ,
\end{equation}
at $m = 1$, $2$, $3$, ... and with
\begin{equation}
\label{eq2b}
\alpha_1 \simeq 1.2\mbox{--}1.3
\ \ \mbox{and} \ \
\alpha_m \simeq \alpha_1 \left[ (2m-1)!! \right]^{1/m}
. 
\end{equation}
\end{subequations}
The scaling is actually not exact at $m \geqslant 2$. From the above relation among the coefficients $\alpha_m$, it follows that the probability density function (PDF) of $u_z$ is closely Gaussian \cite{mm13}. Then, as an idealization, we assume the exact Gaussianity and hence the exact scaling at all the orders $2m$. Its height range is identical to the constant-stress layer, where the scaling of the mean velocity $U_z$ is also logarithmic.

The logarithmic scaling of velocity fluctuations $u_z$ has been predicted by a phenomenological model of energy-containing eddies that are extending from or are attached to the wall, i.e., the attached eddy hypothesis \cite{t76,pc82}. Nevertheless, we would like to derive it from some mathematics of the constant-stress layer. This is because such derivations exist for the scaling of the mean velocity $U_z$ \cite{s48,ll59,my71}. In addition, those mathematics are by themselves of interest. We obtain the necessary and sufficient conditions for the logarithmic scaling of the velocity fluctuations $u_z$. They are compared with characteristics of the attached eddies and of the logarithmic scaling of the mean velocity $U_z$.

\section{Scaling of Mean Velocity} \label{S2}

We summarize one of the mathematical derivations of the logarithmic scaling of the mean velocity $U_z$ \cite{my71}, which is to be applied to velocity fluctuations $u_z$ in Sec. \ref{S3}. The height range of that scaling, i.e., the constant-stress layer, is far from the wall and is also far from the outer region of the turbulence. While $u_{\ast}$ serves as a characteristic velocity that is a constant in units of velocity, there is no characteristic height. The difference of the mean velocity $U_{z_1}-U_{z_2}$ is described by $u_{\ast}$, $z_1$, and $z_2$ alone.

The ratio $(U_{z_1}-U_{z_2})/u_{\ast}$ is thereby invariant under the scale transformation $(x,y,z) \rightarrow ({\mit\Lambda}x,{\mit\Lambda}y,{\mit\Lambda}z)$ with ${\mit \Lambda} > 0$, so long as ${\mit\Lambda} z_1$ and ${\mit\Lambda} z_2$ are left in the constant-stress layer. It is a function of the height ratio $z_1/z_2$ as
\begin{equation}
\label{eq3}
\frac{U_{z_1}-U_{z_2}}{u_{\ast}} = f \left( \frac{z_1}{z_2} \right) .
\end{equation}
From $(U_{z_1}-U_{z_2})+(U_{z_2}-U_{z_3})=U_{z_1}-U_{z_3}$, we obtain $f(z_1/z_2)+f(z_2/z_3)=f(z_1/z_3)$. Its solution is a logarithmic function $f(z_1/z_2) \varpropto \ln (z_1/z_2)$, corresponding to Eq.~(\ref{eq1}).

Thus, the logarithmic scaling is from the existence of the characteristic velocity $u_{\ast}$ and from the nonexistence of any characteristic height in the constant-stress layer. If the characteristic velocity were also not existent, we would instead expect a power-law scaling as in the case of the inertial range of locally homogeneous and isotropic turbulence \cite{ko41,m15}.

An important remark is that the constant-stress layer has no definite boundary \cite{my71}. If there were a definite boundary, it would serve as a characteristic height so that the logarithmic scaling would not exist. The constant-stress layer is actually an asymptotic idealization at high Reynolds numbers, where the wall and the outer region of the turbulence are separated enough. Accordingly asym-ptotic is the logarithmic scaling of the mean velocity $U_z$ and of the velocity fluctuations $u_z$.

\begin{table*}[tbp]
\begingroup
\squeezetable
\caption{\label{t1} Our mathematical conditions compared with the phenomenology of eddies attached to the wall.}
\begin{ruledtabular}
\begin{tabular}{rll}
      & mathematical condition                                                                    & phenomenology \\ \hline
  (i) & The ratio $(u_{z_1} - u_{z_2})/u_{\ast}$ has a PDF that depends only                      & The eddies have a characteristic velocity $u_{\ast}$ but have no\\
      & on $z_1/z_2$.                                                                             & characteristic size. \\
 (ii) & The ratios $(u_{z_1} - u_{z_2})/u_{\ast}$, ..., and $(u_{z_{N-1}} - u_{z_N})/u_{\ast}$ do & The eddies are distributed independently.\\
      & not depend on one another.                                                                & \\
(iii) & The ratio $(u_{z_1} - u_{z_2})/u_{\ast}$ has a self-similar PDF.                          & The eddies have self-similar velocity fields.
\end{tabular}
\end{ruledtabular}
\endgroup
\end{table*}

\section{Scaling of Velocity Fluctuations} \label{S3}

To derive the logarithmic scaling of velocity fluctuations $u_z$, the following conditions are set for the constant-stress layer at some fixed $x$-$y$ position: (i) $(u_{z_1} - u_{z_2})/u_{\ast}$ has a PDF that depends only on $z_1/z_2$ for any pair of $z_1 > z_2$ \cite{my71}; (ii) $(u_{z_1} - u_{z_2})/u_{\ast}$, $(u_{z_2} - u_{z_3})/u_{\ast}$, ..., and $(u_{z_{N-1}} - u_{z_N})/u_{\ast}$ do not depend on one another for any finite series of $z_1 > z_2 > \cdots > z_N$; (iii) $(u_{z_1} - u_{z_2})/u_{\ast}$ has a self-similar PDF, i.e., of the same shape, for all the pairs of $z_1 > z_2$. Here, in order that the separations between the heights $z_{n-1}$ and $z_n$ are larger than the characteristic length $\nu/u_{\ast}$ for the viscosity $\nu$ \cite{my71}, we have set it to be negligibly small. The condition (i) is an extension of that for the scaling of the mean velocity $U_z$.

These are idealized conditions. For example, although our condition (ii) implies that the velocity differences $u_{z_{n-1}} - u_{z_n}$ are not correlated with one another, there should actually exist a correlation. Nevertheless, it is expected to be much weaker than the correlation among the velocities $u_{z_n}$. As described below, such conditions are at the same level of idealizations as for the Gaussianity of $u_z$ in Sec.~\ref{S1} and for the existence of the constant-stress layer in Sec.~\ref{S2}. Beyond these idealizations, any mathematical study would have to be deferred until more details are known (see also Table~\ref{t1}).

Temporarily, we extrapolate the constant-stress layer up to a hypothetical height $z_0$ such that $u_{z_0}$ is equal to $0$. The range from $z_0$ to $z$ is divided as $z_0 > z_1 > \cdots  > z_N = z$ with a constant $\lambda = z_{n-1}/z_n > 1$,
\begin{subequations}
\begin{equation}
\label{eq4a}
\frac{u_z}{u_{\ast}} = \frac{u_{z_N}-u_{z_0}}{u_{\ast}} = \sum_{n=1}^{N} \frac{u_{z_n}-u_{z_{n-1}}}{u_{\ast}}
\ \ \mbox{with} \ \
N = \frac{\ln (z_0 / z)}{\ln \lambda}.
\end{equation}
For these fluctuations, instead of the moments, we study the cumulants \cite{ks77}. That of $u_z$ at the order $\tilde{m}$ is defined as 
\begin{equation}
\label{eq4b}
\langle u_z^{\tilde{m}} \rangle_c = \left. \frac{d^{\tilde{m}}}{d(i \xi)^{\tilde{m}}} \ln \langle \exp (i \xi u_z ) \rangle \right\vert_{\xi=0} .
\end{equation}
The cumulants are yet related with the moments, e.g., $\langle u_z^2 \rangle_c = \langle (u_z-\langle u_z \rangle)^2 \rangle$, $\langle u_z^3 \rangle_c = \langle (u_z-\langle u_z \rangle)^3 \rangle$, and $\langle u_z^4 \rangle_c = \langle (u_z-\langle u_z \rangle)^4 \rangle - 3 \langle (u_z-\langle u_z \rangle)^2 \rangle^2$, albeit with $\langle u_z \rangle_c = \langle u_z \rangle =0$ in the present case. Those at $\tilde{m} \geqslant 3$ are also equal to $0$ if $u_z$ is Gaussian \cite{ks77} as for the scaling in Eq. (\ref{eq2}).

The condition (i) implies that all the summands of Eq. (\ref{eq4a}) have the same PDF as some random variable $r_{\lambda}$, while the condition (ii) implies that they do not depend on one another. Since any cumulant of a sum of independent random variables is the sum of cumulants of those variables \cite{ks77},
\begin{equation}
\label{eq4c}
\frac{\langle u_z^{\tilde{m}} \rangle_c}{u_{\ast}^{\tilde{m}}} 
= \sum_{n=1}^{N} \frac{\langle ( u_{z_n}-u_{z_{n-1}} )^{\tilde{m}} \rangle_c}{u_{\ast}^{\tilde{m}}} 
= \frac{\ln (z_0 / z)}{\ln \lambda} \langle r_{\lambda}^{\tilde{m}} \rangle_c.
\end{equation}
If we rearrange Eq.~(\ref{eq4c}) to separate $\langle r_{\lambda}^{\tilde{m}} \rangle_c / \ln \lambda$, it turns out to be independent of $\lambda$ and to be some constant $\beta_{\tilde{m}}$. Then, the hypothetical height $z_0$ is removed by taking the difference between the heights $z_1$ and $z_2$ as
\begin{equation}
\label{eq4d}
\frac{\langle u_{z_1}^{\tilde{m}} \rangle_c-\langle u_{z_2}^{\tilde{m}} \rangle_c}{u_{\ast}^{\tilde{m}}} 
= -\beta_{\tilde{m}}  \ln \left( \frac{z_1}{z_2} \right).
\end{equation}
\end{subequations}
Especially at $\tilde{m}=2$, Eq.~(\ref{eq4d}) corresponds to the logarithmic scaling of $\langle u_z^{2m} \rangle^{1/m}$ at $m=1$ in Eq.~(\ref{eq2}) via $\beta_2 = \alpha_1$. The scaling at $m \geqslant 2$, corresponding to $\langle u_z^{\tilde{m}} \rangle_c = 0$ and thereby to $\beta_{\tilde{m}} = 0$ at $\tilde{m} \geqslant 3$, is not yet derived from our conditions (i) and (ii) alone.

The condition (iii) implies that the shape of the PDF is identical between the summands and the sum of Eq.~(\ref{eq4a}). Hence, $u_{z_n}-u_{z_{n-1}}$ has to follow a stable distribution \cite{ks77}. If and only if the distribution is stable, the shape of its PDF is invariant under addition of its independent variables. To obtain a finite value for $\langle u_z^2 \rangle_c$, the stable distribution has to be Gaussian. Such a Gaussian PDF yields $\langle u_z^{\tilde{m}} \rangle_c = 0$ at $\tilde{m} \geqslant 3$ \cite{ks77}. From $\langle u_z^{2m} \rangle^{1/m} \varpropto \langle u_z^2 \rangle_c$, we derive the logarithmic scaling of Eq.~(\ref{eq2}).

On the other hand, if Eq.~(\ref{eq2}) holds, it yields our conditions (i)--(iii) via Eq.~(\ref{eq4a}). This is because any Gaussian distribution is reproduced as the distribution of the sum of any number of independent and identically distributed Gaussian random variables \cite{ks77}.

Thus, our conditions (i)--(iii) are necessary and sufficient for the logarithmic scaling of velocity fluctuations $u_z$ in Eq.~(\ref{eq2}). The conditions (i) and (ii) are essential to the existence of the characteristic velocity $u_{\ast}$ and to the nonexistence of any characteristic height, which are in turn essential to the existence of a logarithmic scaling. For example, without the condition (ii), the velocity differences $u_{z_n} - u_{z_{n-1}}$ would exhibit a correlation. Its length scale would characterize the constant-stress layer. To restrict the logarithmic scaling to the form of Eq.~(\ref{eq2}), we have used the condition (iii), where the shape of the PDF of $u_z$ is not characterized by any scale.

\section{Comparison with Phenomenology} \label{S4}

The logarithmic scaling of velocity fluctuations $u_z$ has been predicted by the attached eddy hypothesis \cite{t76,pc82}, which is a phenomenological model of a random superposition of energy-containing eddies that are attached to the wall. An asymptotically infinite number of such eddies are set at the same position and at the same time. While their velocity fields are self-similar to one another with a common characteristic velocity $u_{\ast}$, their sizes are distributed with no characteristic size. The size distribution is accordingly a power law. Its exponent has been determined with conditions of the eddies near the wall so as to reproduce the constant-stress layer.

Since the velocity $u_z$ is due to attached eddies with vertical sizes larger than $z$, the velocity difference $u_{z_1} - u_{z_2}$ is due to those between $z_1$ and $z_2$. Then, as summarized in Table \ref{t1}, the characteristics of the attached eddies correspond closely to the conditions (i)--(iii) of our derivation. It serves as a mathematical explanation for those eddies to reproduce the logarithmic scaling.

The model of the attached eddies has been extended to yield a greater variety of predictions \cite{m01,wm15}. However, especially if such an eddy is a coherent structure, the reason for its existence is not yet known. On the other hand, our mathematical conditions (i)--(iii) are reasonably expected for the constant-stress layer. The pursue of these mathematics is at least equally promising.

\section{Relation Between Coefficients} \label{S5}

The coefficient $\alpha_1$ of the logarithmic scaling of velocity fluctuations $u_z$ is related with the coefficient $1/\kappa$ of the scaling of the mean velocity $U_z$, by using the mathematics described in Sec.~\ref{S3}.

The conditions (i)--(iii) permit us to regard $u_z/u_{\ast}$ as a stochastic Wiener process \cite{f68}, e.g., a Brownian motion, for the time parameter $\tau = \ln (z_0/z)$. Any process $\chi_{\tau}$ at $\tau \geqslant 0$ is a Wiener process if $\chi_{\tau} = 0$ at $\tau=0$, if $\chi_{\tau}$ is a continuous function of $\tau$, if $\chi_{\tau_1} - \chi_{\tau_2}$ has a Gaussian PDF that depends only on $\tau_1 - \tau_2$, and if $\chi_{\tau_1} - \chi_{\tau_2}$, $\chi_{\tau_2} - \chi_{\tau_3}$, ..., and $\chi_{\tau_{N-1}} - \chi_{\tau_N}$ do not depend on one another.

Since any Wiener process is described by a binomial random walk in the limit of its time interval $\delta \tau \rightarrow 0$ \cite{f68}, the velocity fluctuations $u_z$ are described as well. This is the case even if the mean velocity $U_z$ is included as $\chi_{\tau} = (U_z+u_z)/u_{\ast}$. We require two independent parameters to determine the Gaussian distribution of $\chi_{\tau}$. The one is the von K\'arm\'an constant $\kappa$, while the other is set to be some constant $\gamma > 0$. For each interval of the time $\delta \tau$, the displacement $\delta \chi_{\tau} = \chi_{\tau + \delta \tau} - \chi_{\tau}$ has to be either of $\pm (\gamma \delta \tau / \kappa)^{1/2}$ with a different probability $p$ as
\begin{subequations}
\begin{equation}
\label{eq5}
\delta \chi_{\tau}=
  \begin{cases}
  + (\gamma \delta \tau / \kappa)^{1/2} \ \  \mbox{with} \ \  p = 1/2- (\delta \tau /\gamma \kappa)^{1/2}/2, \\  
  - (\gamma \delta \tau / \kappa)^{1/2} \ \  \mbox{with} \ \  p = 1/2+ (\delta \tau /\gamma \kappa)^{1/2}/2.
  \end{cases}
\end{equation}
The result in the limit $\delta \tau \rightarrow 0$ is
\begin{equation}
\label{eq5b}
\langle \delta \chi_{\tau} \rangle = -\frac{\delta \tau}{\kappa}
\ \  \mbox{and} \ \ 
\langle (\delta \chi_{\tau} - \langle \delta \chi_{\tau} \rangle )^2 \rangle = \frac{\gamma \delta \tau}{\kappa}.
\end{equation}
They correspond to the decrease in $U_z/u_{\ast} = \langle \chi_{\tau} \rangle$ and to the increase in $\langle u_z^2 \rangle /u_{\ast}^2 = \langle (\chi_{\tau} - \langle \chi_{\tau} \rangle )^2 \rangle$ with a decrease in $z / z_0 = \exp (-\tau)$. By comparing Eq.~(\ref{eq5b}) with Eqs.~(\ref{eq1}) and (\ref{eq2}) in their limits $z_1 \rightarrow z_2$, 
\begin{equation}
\label{eq5c}
\alpha_1 = \frac{\gamma}{\kappa} .
\end{equation}
\end{subequations}
Thus, for the mathematical consistency, the scaling coefficient $\alpha_1$ for the velocity fluctuations $u_z$ has to be proportional to the scaling coefficient $1/\kappa$ for the mean velocity $U_z$. With $\kappa \simeq 0.4$ and $\gamma \simeq 0.5$, we reproduce the observed value of $\alpha_1 \simeq 1.2$--$1.3$ \cite{hvbs12,mmhs13,mm13,hvbs13,swm14,vhm15}.

The meaning of the parameter $\gamma$ is studied with the budget of the kinetic energy. We use Eq.~(\ref{eq2}) to obtain the difference in the energy of the fluctuations $u_z$ per unit volume between the heights $z$ and $z + \delta z$,
\begin{subequations}
\begin{equation}
\label{eq6a}
-\frac{\rho}{2} \frac{d \langle u_z^2 \rangle}{dz} \delta z = \alpha_1 \frac{\rho u_{\ast}^2}{2} \frac{\delta z}{z}.
\end{equation}
The energy is from the Reynolds stress $\rho \langle -u_z w_z \rangle = \rho u_{\ast}^2$ acting on the mean velocity $U_z$. We use Eq.~(\ref{eq1}) to obtain the energy of the mean velocity at that height $z$ converted per unit volume per unit time to the total energy of the velocity fluctuations $\rho \langle \vert \mbox{\boldmath $u$}_z \vert^2 \rangle /2 = \rho \langle u_z^2 + v_z^2 + w_z^2 \rangle /2$. Here $v_z$ is the spanwise velocity. The result is
\begin{equation}
\label{eq6b}
\rho \langle -u_z w_z \rangle \frac{dU_z}{dz} = \frac{1}{\kappa} \frac{\rho u_{\ast}^3}{z}.
\end{equation}
This is assumed to sustain the local turbulence between the heights $z$ and $z + \delta z$ \cite{t61} over the duration ${\mit \Gamma} \delta z /u_{\ast}$. Here ${\mit \Gamma} >0$ is a constant of about the order of unity. By comparing Eq.~(\ref{eq6a}) with Eq.~(\ref{eq6b}), we obtain a relation equivalent to Eq.~(\ref{eq5c}),
\begin{equation}
\label{eq6c}
\alpha_1 = 2{\mit \Gamma} \frac{\langle u_z^2 \rangle}{\langle \vert \mbox{\boldmath $u$}_z \vert^2 \rangle} \frac{1}{\kappa}.
\end{equation}
\end{subequations}
The parameter $\gamma$ is thus equivalent to $2{\mit \Gamma} \langle u_z^2 \rangle / \langle \vert \mbox{\boldmath $u$}_z \vert^2 \rangle$. It is determined by the local turbulence, where the kinetic energy is redistributed among the streamwise, spanwise, and vertical velocities, is transferred to the smaller length scales, and is dissipated into heat. We consider that $\gamma$ is more fundamental than the coefficient $\alpha_1$.

If the local turbulence in the constant-stress layer is determined by $u_{\ast}$ and $z$ alone, the value of $\gamma$ is universal. This is at least a good approximation, but yet not known is whether this is exactly the case. For example, if the wall surface is rough, it might affect the value of $\langle u_z^2 \rangle / \langle \vert \mbox{\boldmath $u$}_z \vert^2 \rangle$ \cite{ka94,m03}. The experiments and simulations are required much more.

\section{Concluding Remarks} \label{S6}

For wall turbulence, we have derived the logarithmic scaling of velocity fluctuations $u_z$ in Eq.~(\ref{eq2}), by using mathematics of random variables in the constant-stress layer. The existence of the characteristic velocity $u_{\ast}$ and the nonexistence of any characteristic height have yielded the conditions (i)--(iii), which are necessary and sufficient for that scaling. They correspond to the mathematical description of a phenomenological model of energy-containing eddies that are attached to the wall. We have also used those mathematics to relate the scaling coefficient $\alpha_1$ with the von K\'arm\'an constant $\kappa$.

There is another derivation of the logarithmic scaling \cite{h12}, which assumes an overlap between the inner scaling of $\langle u_z^{2m} \rangle^{1/m}$ from near the wall and its outer scaling from the large heights $z$ \cite{i37,m38}. However, any overlap occurs where the inner scaling is asymptotically independent of the wall characteristics and the outer scaling is asymptotically independent of the large-height characteristics. Only the friction velocity $u_{\ast}$ is left as the characteristic scale. That assumption of the existence of an overlap  of $\langle u_z^{2m} \rangle^{1/m}$ is thereby equivalent to assuming the existence of its logarithmic scaling. We instead have to consider the fluctuations $u_z$ themselves.

The idealized Gaussianity of the velocity fluctuations $u_z$ is usually explained by applying the central limit theorem to their Fourier transforms \cite{m03,b53,m02} or to the velocities of the individual attached eddies \cite{mm13}. It is such that a sum of random variables becomes Gaussian with an increase in their total number \cite{ks77}. This theorem is applicable to the sum of $(u_{z_n}-u_{z_{n-1}})/u_{\ast}$ in Eq.~(\ref{eq4a}). However, the Gaussianity is obtained only asymptotically in the limit $N \rightarrow +\infty$, i.e., $z/z_0 \rightarrow 0$. Then, the coefficients $\beta_{\tilde{m}}$ at $\tilde{m} \geqslant 3$ in Eq.~(\ref{eq4d}) differ from the Gaussian value of $0$. Although this could be the case if the PDF were asymptotically determined during the mean momentum transfer to the lower heights, the momentum is transferred locally and instantaneously to the larger heights as well as to the smaller heights. Their velocity fluctuations interact with one another and should have settled into some self-similar state as assumed in our condition (iii).

From the phenomenological model of the attached eddies, it is also expected that the spanwise velocity $v_z$ has $(\langle v_{z_1}^{2m} \rangle^{1/m} -\langle v_{z_2}^{2m} \rangle^{1/m}) / u_{\ast}^2 \varpropto \ln (z_1/z_2)$ while the vertical velocity $w_z$ has $(\langle w_{z_1}^{2m} \rangle^{1/m} -\langle w_{z_2}^{2m} \rangle^{1/m}) / u_{\ast}^2 \simeq 0$ or $\langle w_z^{2m} \rangle^{1/m}/ u_{\ast}^2 \simeq \mbox{constant}$ \cite{t76,pc82,wm15}. This is reasonable if $v_z$ and $w_z$ satisfy our conditions (i)--(iii). Here $w_z$ is regarded as a special case where the coefficient of its logarithmic scaling is close to $0$ as a result of the blocking by the wall. The coefficient would not be exactly equal to $0$. Otherwise, $\langle u_z^2 \rangle / \langle \vert \mbox{\boldmath $u$}_z \vert^2 \rangle$ would not be a constant within the constant-stress layer, and hence we would not be permitted to use Eq.~(\ref{eq6c}) for the coefficient $\alpha_1$ of the streamwise velocity $u_z$.

The logarithmic scaling is expected for fluctuations of any other random variable, e.g., the density of a passive admixture in the constant-stress layer \cite{ll59,my71}, if it satisfies our necessary and sufficient conditions (i)--(iii). This is not restricted among those of the wall turbulence. For a variety of fields, it would be of great interest to search for such logarithmic scaling of random variables.

\begin{acknowledgments}
This work was supported in part by KAKENHI Grant No. 25340018.
\end{acknowledgments}

\end{document}